\documentclass[12pt]{article}

\usepackage{comment}
\usepackage{tikz}
\usepackage{caption}

\usepackage{mathtools}

\usepackage[mathscr]{euscript}

\usepackage{slashed}

\usepackage{amsmath}

\date{}

\title{\textbf{Improved Gauge-Unfixing Formalism through a Prototypical Second-Class System 
}}
\author{ \textbf{J. Ananias Neto$^{a}$,
W. D. Morais$^{b}$ and
R. Thibes$^{c}$}
\\\\
\textit{$^{a}$\small{Instituto de Ci\^encias Exatas}}\\
\textit{\small{Universidade Federal de Juiz de Fora}}\\
\textit{\small{Rua José Lourenço Kelmer
São Pedro, Juiz de Fora
-- 36036-330, Brazil}}\\
\textit{$^{b}$\small{Centro Brasileiro de Pesquisas Físicas}}\\
\textit{\small{Rua Dr. Xavier Sigaud 150, Urca,}}\\
\textit{\small{Rio de Janeiro -- 22290-180, Brazil}}\\
\textit{$^{c}$\small{Departamento de Ci\^encias Exatas e Naturais}}\\
\textit{\small{Universidade Estadual do Sudoeste da Bahia}}\\
\textit{\small{Rodovia BR 415, Km 03, S/N, Itapetinga -- 45700-000, Brazil}}
 }

\usepackage{graphicx}
\usepackage{amssymb,amsfonts,amssymb,amsthm}
\usepackage{amsmath}

\begin{document}

\maketitle

\abstract{We contextualize the improved gauge-unfixing (GU) formalism within a rather general prototypical second-class system, obtaining a corresponding first-class equivalent description enjoying gauge invariance which can be applied to several situations.  The prototypical system is chosen to represent a considerable class of relevant models in field theory.  By considering the improved version of the GU formalism, we  show that any gauge-invariant function can be obtained in terms of a specific deformation in phase space, benefiting thus from the fact that no auxiliary variables are needed in the process.  In this way, the resulting converted first-class system is constructed out of the same original canonical variables, preserving the number of degrees of freedom.  We illustrate the technique with an application to the nonlinear sigma model.
}

\section{Introduction}
In general terms, gauge symmetry in physics represents an important and welcome feature in various aspects \cite{Henneaux:1992ig, Jackson:2001ia, Thibes:2020jfp}.   Besides its natural aesthetic appeal, gauge invariance allows a consistent explicit Lorentz covariant description of nature in terms of fundamental symmetries preserving models.  For practical calculations of predictive results, one has an enormous freedom concerning alternative equivalent gauge-fixing choices, being able to pick the most convenient and suitable one for the particular problem at hand.  The functional quantization of gauge-invariant systems can be approached by different well-established systematic formalisms, involving (anti)commuting generalized Faddeev-Popov ghost fields \cite{Faddeev:1967fc, Teitelboim:1987rs} connected with the initial classical gauge transformations, giving rise to forms of BRST symmetry \cite{Becchi:1974xu, Becchi:1974md, Tyutin:1975qk, Becchi:1975nq, Mandal:2022zil} as a corresponding quantum counterpart.  Those facts can be immediately contrasted to the cumbersome Dirac brackets technique \cite{Dirac:1950pj, Anderson:1951ta, Dirac, Sundermeyer:1982gv}, following the opposite path of eliminating redundant variables at the cost of breaking natural symmetries, often creating insurmountable calculating difficulties.  Furthermore, a fair understanding of the interplay between gauge-invariant and non-gauge-invariant descriptions of specific models provides relevant tools for field theory, as can be attested, for instance, when we consider that classical symmetries may be broken at quantum level, eventually leading to anomalous quantum field theories.   From a constrained Hamiltonian viewpoint \cite{Henneaux:1992ig, Thibes:2020jfp, Sundermeyer:1982gv, Gitman:1990qh}, gauge-invariance is related to first class constraints.  In this context, the idea of converting second class constraints to first class ones, in order to generate gauge invariance, appears rather naturally  -- indeed, after first isolated attempts in that direction by means of adding new auxiliary variables to phase space  \cite{Stueckelberg:1957zz, Wess:1971yu},  the physics literature has witnessed the appearance of a great deal of gauge symmetry generating proposals arising from or related to that conversion.     From those, we mention here two consistent 
solid approaches: the BFFT formalism \cite{Batalin:1986aq, Batalin:1986fm, Egorian:1988ss, Batalin:1989dm, Batalin:1991jm}, relying on the insertion of new canonical variables in phase space and the GU formalism \cite{Mitra:1990mp, Anishetty:1992yk, Vytheeswaran:1994np, Neto:2006gt}, which identifies part of the second class constraints as actual gauge-fixing conditions for a parent first class system.  In the present text, we shall be concerned with the latter -- the GU formalism --, applying its improved version to a prototypical second class system encompassing a large class of both mechanical and field theory models.

The first ideas of the GU formalism can be traced back to the early works \cite{Mitra:1989fg, Harada:1988aj} in which gauge invariance is produced without the introduction of auxiliary fields.   The anomalous chiral Schwinger model has been discussed along those lines in \cite{Mitra:1989fg}. In that model, due to the anomaly, gauge symmetry is broken at quantum level. Aiming to circumvent the situation, the corresponding bosonized second-class model can be treated and properly converted to a first-class one, interpreting two of the constraints as gauge-fixing conditions \cite{Mitra:1989fg}.  In reference \cite{Harada:1988aj}, the general GU problem has been clearly stated and discussed for the first time in a model independent way.  Namely, given a non-gauge-invariant theory with second class constraints, the GU approach searches for an equivalent first class one containing only half of the original constraints, characterized now as first class, and in which the remaining old constraints are not seen as natural constraints anymore, but rather possible gauge-fixing choices  \cite{Harada:1988aj}.  In this way, the now first class constraints generate gauge symmetries.  On account of this, the original Hamiltonian must be modified to produce a corresponding first-class one.  In reference \cite{Mitra:1990mp}, we have a nice account of the GU problem in terms of  infinite series expansion of phase space functions, considering the different constraints chains generated by the primary constraints.  A projection operator acting on phase space is introduced in references \cite{Anishetty:1992yk, Vytheeswaran:1994np} to tackle the modified Hamiltonian and other gauge invariant functions, with a corresponding direct application to the chiral Schwinger model in \cite{Vytheeswaran:1992ph}.  Interesting comparisons between the GU and BFFT formalisms can be seen in \cite{Vytheeswaran:2000ug, Ebrahimi:2014mna}.  Certainly one of the main drawbacks of the BFFT conversional approach is the introduction of too many auxiliary variables, one for each first-class-to-be constraint.  Addressing that issue, in reference \cite{Neto:2004dx},  a mixed use of both BFFT and GU methods in the same problem is proposed in which, after an application of the BFFT procedure, arguments from the GU formalism can be used to get rid of auxiliary variables.   An improved version of the GU method is described in reference \cite{Neto:2006gt} in which one first calculates a transformed gauge-invariant form for each of the phase space variables.  Then this gauge-invariant form can be used to easily obtain the modified Hamiltonian, constraints and all gauge invariant phase space functions.  Modern relevant applications of the GU formalism can be seen in references \cite{Vytheeswaran:1997jr, Monemzadeh:2014wma, Sararu:2014sua, Alesci:2018ewg, Alves:2020pty, Alves:2022qts}.

In this letter, we contextualize the improved GU formalism within a propotypical second-class system possessing sufficient generality to describe a considerable class of important both mechanical and field theory physical models.  The model is constructed from an initial set of $M$ second-class constraints $T_\alpha$ satisfying a specific internal consistency condition, which are shown to generate gauge tranformations in the corresponding parent unfixed counterpart.  The constraint functions $T_\alpha$ are introduced by means of a set of corresponding Lagrange multipliers $l^\alpha$, mimicking a common situation in various well-known field theory models, regardless of their interpretation as independent variables or phase space functions.  A methodical application of the Dirac-Bermann algotithm produces $4M$ constraints, necessary for time conservation consistency.  Among the $4M$ Dirac-Bergmann constraints, we identify $2M$ physical ones, which coincide with the so-called {\it true constraints} within the 
Faddeev-Jackiw-Barcelos-Wotzasek (FJBW) sympletic approach \cite{Faddeev:1988qp,
BarcelosNeto:1991kw, Wotzasek:1992wf, Wotzasek:1994ck}.  We show that the $2M$ true constraints involve only the physical variables and allow for a neat successful direct implementation of the GU method.  Thus, it is clear that second-class physical systems falling into this prototypical category can immediately benefit from a fresh gauge-invariant description, in quite a general framework.  
We illustrate our results in field theory within the nonlinear sigma model, which can be treated as a particular case of our prototypical second-class system. 

Our work is organized as follows.  In Section {\bf 2} below, we introduce a prototypical second-class system and discuss its canonical constraints structure.  A family of $4M$ second class constraints is identified, out of which only $2M$ ones are precisely characterized as true constraints, according to the FJBW formalism.  In Section {\bf 3}, we apply the improved GU formalism to the prototypical second-class system, obtaining the deformed new variables in phase space.  This allows us to further generate a parent first-class system, which can be connected to the original second-class one by a suitable gauge-fixing.  In Section {\bf 4}, we illustrate our general ideas in the non-linear sigma model.  We end in Section {\bf 5} with some concluding final remarks.

\section{A Prototypical Second-Class System}
Second-class systems are Dirac-Bergmann constrained Hamiltonian dynamical systems \cite{Dirac:1950pj, Anderson:1951ta, Dirac} with an invertible Poisson brackets (PB) constraints matrix. Thus,  as a direct consequence from its very defining property, second-class systems do not enjoy gauge freedom.  The main claim of the GU formalism then states that those second-class systems can be thought of as coming from an equivalent first-class one fixed in a specific gauge.  To set up matters, notation and conventions, let us consider a prototypical second-class system defined by a Lagrangian function of the form \cite{Pandey:2021myh}
\begin{equation}\label{L}
L(l^\alpha,q^k,{\dot{q}}^k)=\frac{1}{2}f_{ij}(q^k){{\dot{q}}^i}{{\dot{q}}^j}
-V(q^k)
-l^\alpha T_\alpha(q^k)
\,,
\end{equation}
living in a configuration space constructed from the generalized coordinates
$q^i, l^\alpha$, with Latin and Greek indexes running respectively through $i, j, k = 1, \dots, N$, and $\alpha,\beta,\gamma=1,\dots,M$.  Further, $f_{ij}(q^k)$ denotes a non-degenerated symmetric tensor, $V(q^k)$
a twice differentiable real function representing an arbitrary potential and $T_\alpha(q^k)$ stands for a set of $M$
 thrice differentiable real functions playing the role of natural constraints coming from the $l^\alpha$ equations of motion.   The singularity of (\ref{L}) as a Dirac-Bergmann constrained system comes from the fact that Lagrange multipliers time derivatives $\dot{l}^\alpha$ do not explicitly show up and, consequently, their canonical momenta are degenerate.

By introducing the upper-index quantity $f^{ij}(q^k)$, standing for the inverse of $f_{ij}$ satisfying
\begin{equation}
f^{\,ik}\,f_{kj} = f_{jk}\,f^{\,ki} = \delta^{\,i}_{\,j}\,,
\end{equation}
we may compute the canonical Hamiltonian associated to (\ref{L}) as
\begin{equation}\label{H}
H=\frac{1}{2}f^{ij}(q^k)p_ip_j+V(q^k)+l^\alpha T_\alpha(q^k)
\,,
\end{equation}
where we have introduced the canonical momenta
\begin{equation}
p_i\equiv \frac{\partial L}{\partial {\dot{q}}^i}\,\, 
\mbox{ and }\,\,
\pi_\alpha \equiv \frac{\partial L}{\partial {\dot{l}}^\alpha}
\,.
\end{equation}
In fact, since $\pi_\alpha=0$, the Legendre transformation leading to (\ref{H}) is singular and is actually well-defined only on the $(2N+M)$-dimensional hypersurface of primary constraints 
\begin{equation}\label{Xi0}
\chi_{(0)\alpha} \equiv \pi_\alpha\,,\,\,\,\,\alpha=1,\dots,M
\,.
\end{equation}
Following blindly the DB algorithm \cite{Dirac:1950pj, Anderson:1951ta, Dirac}, we impose the stability of the constraints under time evolution.   As a result, additionally to (\ref{Xi0}), three more constraint families are generated,
\begin{equation}\label{Xi1Xi2}
\chi_{(1)\alpha}=T_\alpha
\,,~~~~~~
\chi_{(2)\alpha} = f^{ij}p_iT_{\alpha, j}
\,,
\end{equation}
and
\begin{equation}\label{Xi3}
\chi_{(3)\alpha} = \frac{1}{2}Q_\alpha^{ij}p_ip_j
-v_\alpha-l^\beta w_{\alpha\beta}
\,,
\end{equation}
always with $\alpha=1,\dots,M$, constituting a total of $4M$ constraints in phase space.
For notation convenience, in the RHS of equation (\ref{Xi3}) we have introduced the $q^k$-dependent
quantities $Q^{ij}_\alpha$, $v_\alpha$ and $w_{\alpha \beta}$  respectively defined as
\begin{equation}\label{Qij}
Q^{ij}_\alpha\equiv
{\left(f^{il}T_{\alpha, l}\right)}_{,k} f^{kj}
+{\left(f^{jl}T_{\alpha, l}\right)}_{,k} f^{ki}
-f^{kl}T_{\alpha, k} f^{ij}_{\,\,\,\,,l}
\,,
\end{equation}
\begin{equation}\label{v}
v_\alpha\equiv f^{ij}T_{\alpha, i}V_j
\,,
\end{equation}
and
\begin{equation}\label{w}
w_{\alpha\beta}\equiv f^{ij}T_{\alpha, i} T_{\beta, j}
\,.
\end{equation}
We use commas to denote partial derivatives with respect to the generalized variables $q^i$, for instance
\begin{equation}
T_{\alpha,i}\equiv \frac{\partial T_\alpha}{\partial {{q}}^i}\,,~~~~~~
f^{ij}_{\,\,\,\,,l} \equiv \frac{\partial f^{ij}}{\partial {{q}}^l}
\,,
\end{equation}
and so on and so forth. 
The key condition for the prototypical system (\ref{L}) represent a second-class system {\it de facto}, is the invertibility of the above matrix (\ref{w}), which here by all means we definitely assume to be true via
\begin{equation}\label{assumption}
w\equiv \det w_{\alpha\beta} \neq 0
\,.
\end{equation}
This is necessary and sufficient to guarantee the second-class nature for 
the whole set of constraints $\chi_{(r)\alpha}$ with $r=0,\dots,3$, as can be seen by
computing their Poisson brackets and writing the resulting PB constraint matrix as 
\begin{equation}\label{CM}
\Delta_{(rs)\alpha\beta}\equiv\lbrace\chi_{(r)\alpha}, \chi_{(s)\beta} \rbrace =
\displaystyle
\left[
\begin{array}{cccc}
0 & 0 & 0 & w_{\alpha\beta} \\ 
0 & 0 &w_{\alpha\beta} &E_{\alpha\beta} \\
0 & -w_{\alpha\beta}  & M_{\alpha\beta} & R_{\alpha\beta} \\
-w_{\alpha\beta}   & -E_{\beta\alpha} & -R_{\beta\alpha} & N_{\alpha\beta}
\end{array}
\right]
\,,
\end{equation}
for $r,s=0,\dots,3$, with the short-hand conventions
\begin{equation}
M_{\alpha\beta}\equiv
\lbrace
\chi_{(2)\alpha}, \chi_{(2)\beta}
\rbrace
\,,
\end{equation}
\begin{equation}
E_{\alpha\beta}\equiv
\lbrace
\chi_{(1)\alpha}, \chi_{(3)\beta}
\rbrace
\,,\,\,\,\,
R_{\alpha\beta}\equiv
\lbrace
\chi_{(2)\alpha}, \chi_{(3)\beta}
\rbrace
\,,
\end{equation}
and
\begin{equation}
N_{\alpha\beta}\equiv
\lbrace
\chi_{(3)\alpha}, \chi_{(3)\beta}
\rbrace
\,.
\end{equation}
In fact, the determinant of the constraint matrix (\ref{CM}) depends only on its secondary diagonal and is given by
\begin{equation}\label{w4}
\det \Delta_{(rs)\alpha\beta} =
w^4
\end{equation}
which clearly shows that, under the assumption (\ref{assumption}), the prototypical system (\ref{L}) is indeed second-class.

Although the remaining further entries below the main diagonal in the constraint matrix (\ref{CM}) do not affect its determinant value in (\ref{w4}), we compute them explicitly for completeness and future reference as
\begin{equation}\label{M}
M_{\alpha\beta}=f^{ij}p_k
\left[
{\left(f^{kl}T_{\alpha ,l}\right)}_{,i}T_{\beta ,j}
-
{\left(f^{kl}T_{\beta ,l}\right)}_{,i}T_{\alpha ,j}
\right]
=
p_k M^k_{\alpha\beta}
\,,
\end{equation}

\begin{equation}\label{DR}
E_{\alpha\beta}=T_{\alpha ,i}Q^{ij}_\beta p_j
\,,
~~~
R_{\alpha\beta}
=
p_i p_j R^{ij}_{\alpha\beta}+ f^{ij}T_{\alpha ,i} v_{\beta ,j} 
+ l^\gamma f^{ij}T_{\alpha ,i} w_{\beta\gamma ,j}
\,,
\end{equation}
and
\begin{equation}\label{N}
N_{\alpha\beta}= p_i
\left[
\chi_{(3)\alpha,j}Q^{ij}_\beta
-
\chi_{(3)\beta,j}Q^{ij}_\alpha
\right]
=
p_i p_j p_k Q^{ijk}_{\alpha\beta}
+p_i V^i_{\alpha\beta}+p_il^\gamma N^i_{\alpha\beta\gamma}
\,,
\end{equation}
with the definitions
\begin{equation}\label{D1}
M^k_{\alpha\beta}\equiv f^{ij}
\left[
\left(f^{kl}T_{\alpha, l}\right)_{,i}T_{\beta, j}
-
\left(f^{kl}T_{\beta, l}\right)_{,i}T_{\alpha, j}
\right]
\,,
\end{equation}
\begin{equation}
Q^{ijk}_{\alpha\beta}\equiv\frac{1}{2}
\left(
Q^{ij}_{\alpha,l}Q^{kl}_\beta
-
Q^{ij}_{\beta,l}Q^{kl}_\alpha
\right)
\,,
\end{equation}
\begin{equation}
V^i_{\alpha\beta}\equiv
v_{\alpha ,j} Q^{ij}_\beta
-
v_{\beta ,j} Q^{ij}_\alpha
\,,~~~~~~
N^i_{\alpha\beta\gamma}\equiv
w_{\alpha\gamma ,j} Q^{ij}_\beta
-
w_{\beta\gamma ,j} Q^{ij}_\alpha
\,,
\end{equation}
and
\begin{equation}\label{D5}
R^{ij}_{\alpha\beta}\equiv
{\left(f^{ik}T_{\alpha ,k}\right)}_{,l}Q^{lj}_\beta
-\frac{1}{2}f^{kl}T_{\alpha ,k} Q^{ij}_{\beta ,l}
\,.
\end{equation}
Similarly to (\ref{Qij}) to (\ref{w}), the quantities in equations (\ref{D1}) to (\ref{D5}) above depend only on the variables $q^k$.

In this way, we have succeeded in pursuing the constraint structure analysis and classification for the canonical Hamiltonian form of the singular prototypical system (\ref{L}) in terms of the standard Dirac-Bergmann algorithm.  However, if we take a closer careful look at the constraint equations (\ref{Xi0}) to (\ref{Xi3}) and its corresponding PB matrix (\ref{CM}), we see that blindly following the canonical Dirac-Bergmann prescription has actually lead us to a unnecessary overkill calculating mountain.  As we have $4M$ second-class constraints, the total number of degrees of freedom of (\ref{L}) is given by
\begin{equation}
\text{DOF}=\frac{2\times(N+M)-4\times M}{2}= N-M\,.
\end{equation}
Naturally, this corresponds to the subtraction of $M$ holonomic constraint conditions $\chi_{(1)\alpha}=T_\alpha$ from $N$ physical variables $q^i$.  It is clear that the Lagrange multiplier variables $l^\alpha$ along with their canonical momenta $\pi_\alpha$ do not represent physical variables.  Thus the relevant constraints should not contain the phase space variables $(l^\alpha, \pi_\alpha)$ and are actually given by the $2M$ relations (\ref{Xi1Xi2}) corresponding to $\chi_{(1)\alpha}$ and $\chi_{(2)\alpha}$.  This fact may be more rigorously confirmed by means of the  FJBW symplectic algorithm \cite{Faddeev:1988qp,
BarcelosNeto:1991kw, Wotzasek:1992wf, Wotzasek:1994ck} which rewrites the Lagrangian function (\ref{L}) in first order and detects only the so-called {\it true constraints} which can be checked to coincide exactly with $\chi_{(1)\alpha}$ and $\chi_{(2)\alpha}$ in equation (\ref{Xi1Xi2}).
Along this line, disregarding the auxiliary variables $(l^\alpha, \pi_\alpha)$, we write down the physical second-class Hamiltonian function as
\begin{equation}\label{Hsc}
H_{sc}=H_0+l^\alpha(q^k,p_k) T_\alpha(q^k)
\,,
\end{equation}
with
\begin{equation}\label{H0}
H_0\equiv
\frac{1}{2}f^{ij}(q^k)p_ip_j+V(q^k)
\end{equation}
and $l^\alpha=l^\alpha(q^k,p_k)$ now obtained from (\ref{Xi3}) as
\begin{equation}
l^\alpha(q^k,p_k) = 
 \frac{1}{2}w^{\alpha\beta}(q^k)Q_\beta^{ij}(q^k)p_ip_j
-w^{\alpha\beta}(q^k)v_\beta(q^k)
\,,
\end{equation}
with $Q_\alpha^{ij}(q^k)$ and $v_\alpha(q^k)$ still given by equations (\ref{Qij}) and (\ref{v}).

Summarizing, the singular Lagrangian function (\ref{L}) leads to a consistent second-class Hamiltonian system described by equation (\ref{Hsc}) along with $2M$ second-class constraints (\ref{Xi1Xi2}) and PB algebra given by
\begin{equation}\label{SCS1}
\lbrace
\chi_{(1)\alpha}, \chi_{(1)\beta}
\rbrace
=0\,,
~~~~
\lbrace
\chi_{(2)\alpha}, \chi_{(2)\beta}
\rbrace
= M_{\alpha\beta}
\,,
~~~~
\lbrace
\chi_{(1)\alpha}, \chi_{(2)\beta}
\rbrace
=w_{\alpha\beta}
\,,
\end{equation}
with $M_{\alpha\beta}$ and $w_{\alpha\beta}$ given by (\ref{M}) and (\ref{w}), and
\begin{equation}\label{SCS3}
\lbrace
H_{sc}, \chi_{(1)\alpha}
\rbrace
= A_\alpha^{\,\,\beta} \chi_{(1)\beta} + B_\alpha^{\,\,\beta}\chi_{(2)\beta}
\,,
\end{equation}
\begin{equation}\label{SCS4}
\lbrace
H_{sc}, \chi_{(2)\alpha}
\rbrace
= C_\alpha^{\,\,\beta} \chi_{(1)\beta} + D_\alpha^{\,\,\beta}\chi_{(2)\beta}
\,,
\end{equation}
with just introduced phase space functions given by
\begin{equation}\label{AB}
A_\alpha^{\,\,\beta} = -w^{\beta\gamma}Q^{ij}_\gamma T_{\alpha, i} p_j \,,~~~~ B_\alpha^{\,\,\beta} = -\delta_\alpha^{\,\,\beta} \,,
\end{equation}
\begin{equation}\label{CD}
C_\alpha^{\,\,\beta} = f^{ij}T_{\alpha ,j} l^\beta_i - w^{\beta\gamma}Q^{kl}_\gamma (f^{ij}T_{\alpha ,j})_{,k} p_i p_l ~~~\mbox{ and }~~~ D_\alpha^{\,\,\beta} = 0\,.
\end{equation}

As we discussed in the Introduction, now comes the idea of the GU formalism, which is to consider the second-class system defined by (\ref{SCS1}) to (\ref{SCS4}) as coming from an equivalent parent first-class one complemented with additional gauge-fixing subsidiary conditions.
We shall elaborate further this idea in the next section.

\section{Improved Gauge-Unfixing Formalism}
In conformity with the well-established GU formalism, in this section we construct a new gauge-invariant parent model, related to the second-class system characterized by equations (\ref{Hsc}) to (\ref{SCS4}), in which only half from the $2M$ true constraints (\ref{Xi1Xi2}) shall be identified as actually originally first-class -- the remaining half representing possible subsidiary gauge-fixing conditions.  As discussed in the last section, the $M$ statements $T_\alpha(q^k)=0$ can be naturally regarded as the main holonomic conditions under which the system dynamically evolves.  In this way, from now on we consider
\begin{equation}
\chi_{(1)\alpha}=T_\alpha\,,\,\,\,\,\alpha=1,\dots,M\,,
\end{equation}
as the $M$ first-class constraints responsible for generating continuous $\epsilon^\alpha$-pa\-ra\-m\-e\-trized infinitesimal gauge transformations in arbitrary phase space functions $F(q^k, p_k)$ through
\begin{equation}\label{Fgaugetransformation}
\delta F(q^k, p_k) = \epsilon^\alpha \left\{ F(q^k, p_k)  ,  T_\alpha \right\}
\,,
\end{equation}
within the framework of the sought-for gauge-invariant parent system.
For this assertion to be completely true, we have to replace the original second-class Hamiltonian function (\ref{Hsc}) by a suitable equivalent first-class one $H_{fc}$ and, accordingly, modify the set of relations (\ref{SCS1}) to (\ref{SCS4}) to
\begin{equation}\label{FC1}
\lbrace
T_{\alpha}, T_{\beta}
\rbrace
=0
\,,
\end{equation}
and
\begin{equation}\label{FC2}
\lbrace
T_{\alpha}, H_{fc}
\rbrace
=G_\alpha^{~\beta}T_\beta
\,,
\end{equation}
for some phase space functions $G_\alpha^{~\beta}(q^k, p_k)$.  Note that equation (\ref{FC1}) is already satisfied and we do not need to modify the constraints $T_\alpha$.
In order to calculate $H_{fc}$ in equation (\ref{FC2}), as well as other relevant gauge-invariant quantities,
instead of using the GU method in its originally conceived form \cite{Mitra:1990mp, Anishetty:1992yk, Vytheeswaran:1994np}, we employ the much easier equivalent improved GU approach \cite{Neto:2006gt} in which we start by obtaining the transformed gauge-invariant tilde variables $(\tilde{q}^i, \tilde{p}_i)$.  The latter are defined as deformations
\begin{equation}\label{tildevars}
\begin{aligned}
\tilde{q}^i=\tilde{q}^i(q^k, p_k)\,,
\\
\tilde{p}_i=\tilde{p}_i(q^k, p_k)\,,
\end{aligned}
\end{equation}
satisfying the gauge invariance and boundary conditions respectively given by
\begin{equation}\label{GI}
 \epsilon^\alpha \left\{\tilde{q}^i (q^k, p_k) ,  T_\alpha \right\}
 =  \epsilon^\alpha \left\{\tilde{p}_i (q^k, p_k) ,  T_\alpha \right\}
 =0
\,,
\end{equation}
and 
\begin{equation}\label{B=0}
\begin{aligned}
\tilde{q}^i(q^k, p_k) \vert_{\chi_\alpha = 0} = {q}^i \,,
\\
\tilde{p}_i(q^k, p_k) \vert_{\chi_\alpha = 0} = {p}_i
\,,
\end{aligned}
\end{equation}
where, for simplicity, we have dropped the subscript ${}_{(2)}$ and denote now by $\chi_\alpha\equiv\chi_{(2)\alpha}$ the second half of equations (\ref{Xi1Xi2}).
In fact, (\ref{GI}) demands gauge invariance under $T_\alpha(q^k)$-generated transformations (\ref{Fgaugetransformation}), while (\ref{B=0}) assures that the gauge-fixing 
choice $\chi_\alpha=0$ recovers the original second-class system.  In this way, following \cite{Neto:2006gt}, we expand (\ref{tildevars}) in powers of $\chi_\alpha$ and write
\begin{equation}
\begin{aligned}
\tilde{q}^i=q^i+b^{i\alpha}\chi_\alpha + b^{i\alpha\beta}\chi_\alpha \chi_\beta 
+ b^{i\alpha\beta\gamma}\chi_\alpha \chi_\beta \chi_\gamma 
+~\dots
\,,\\
\tilde{p}_i= p_i+b_i^\alpha \chi_\alpha + b_{i}^{\alpha\beta}\chi_\alpha \chi_\beta 
+ b_{i}^{\alpha\beta\gamma}\chi_\alpha \chi_\beta \chi_\gamma + ~\dots ~
\,,
\end{aligned}
\end{equation}
or equivalently, using a more compact double repeated index sum notation \cite{Thibes:2020bkk},
\begin{equation}\label{qptilde}
\tilde{q}^i= 
b^{i\alpha_{(n)}}\chi_{\alpha_{(n)}}
\,,~~~~
\tilde{p}_i= 
b_i^{\,\,\alpha_{(n)}}\chi_{\alpha_{(n)}}
\,,~~~~ n=0,1,\dots\,,
\end{equation}
with coefficients $b^{i\alpha_{(n)}}$ and $b_i^{\,\,\alpha_{(n)}}$ to be determined.
By plugging (\ref{qptilde}) into (\ref{GI}), we obtain
\begin{equation}\label{deformation}
\tilde{q}^i=q^i ~~\text{ and }~~
\tilde{p}_i= p_i - w^{\alpha\beta}T_{\alpha ,i} \chi_\beta
\,,
\end{equation}
characterizing a $\chi_\alpha$ linear deformation for $p_i$.  Note that $\tilde{p}_i$ can also be written as
\begin{equation}
\tilde{p}_i= {\cal O}_i^{~j} p_j
\,,
\end{equation}
where ${\cal O}_i^{~j}$ denotes a projection operator given by
\begin{equation}\label{O}
{\cal O}_i^{~j} = \delta_i^{\,j}-w^{\alpha\beta}f^{jk}T_{\alpha,i}T_{\beta,k}
\,,
\end{equation}
satisfying
\begin{equation}
{\cal O}_i^{~k} {\cal O}_k^{~j} = {\cal O}_i^{~j}
\,.
\end{equation}
Finally, we may define the first-class Hamiltonian as
\begin{equation}
H_{fc}\equiv H_{sc} (\tilde{q}^k, \tilde{p}_k)
\,,
\end{equation}
and, upon substituting (\ref{deformation}) into (\ref{Hsc}), explicitly obtain
\begin{equation}\label{Hfc}
H_{fc}=\frac{1}{2} f^{ij} p_i p_j - \frac{1}{2} w^{\alpha\beta} \chi_\alpha \chi_\beta + V(q^k) + l^\alpha(\tilde{q}^k,\tilde{p}_k) T_\alpha(q^k)
\,,
\end{equation}
with
\begin{equation}
l^\alpha(\tilde{q}^k,\tilde{p}_k) = \frac{1}{2}w^{\alpha\beta}Q^{ij}_\beta
\left[
p_ip_j
-2w^{\gamma\delta}T_{\gamma,i}p_j\chi_\delta
+\big(w^{\gamma\delta}T_{\gamma,i}\chi_\delta\big)
\big(w^{\epsilon\eta}T_{\epsilon,j}\chi_\eta\big)
\right]
-w^{\alpha\beta}v_\beta
\,.
\end{equation}
It is straightforward now to 
check that $H_{fc}$ indeed satisfies (\ref{FC2}) with
\begin{equation}\label{G=0}
G_\alpha^{~\beta} = 0
\,,
\end{equation} 
showing that we have in fact achieved a parent first-class Abelian strongly involutive description for the original prototypical starting system. 
Actually, the result (\ref{G=0}) was naturally expected and confirms consistency, as we have been able to solve ({\ref{tildevars}) as a strong equality in whole phase space.
To see how all this works in practical terms, in the next section we discuss an application to the nonlinear sigma model.

\section{The Nonlinear Sigma Model}
In this section, we illustrate the previous ideas within the scope of a specific field theory model described by the prototypical second-class system (\ref{L}).  The $O(\mbox{N})$ nonlinear sigma model in a $D$-dimensional Minkowski flat space $x=(x^\mu)$ can be defined by the Lagrangian density function
\begin{equation}\label{ONL}
{\cal L} = \frac{1}{2p}\partial_\mu\phi^a\partial^\mu\phi^a-\frac{\varphi}{2}(\phi^a\phi^a-q^2)
\,,
\end{equation}
with $\phi^a$, $a=1,\dots,\mbox{N}$, denoting a multiplet of real fields, ${\varphi}$ an additional singlet and $p$ and $q$ two given real parameters.\footnote{The natural number N related to the $O(\mbox{N})$ symmetry group should not be confused with the total number of physical coordinates $N$ introduced in Section {\bf 2}.}   As usual, the Lorentz space-time indexes run through $\mu,\nu=0,\dots,D-1$, space indexes through $i,j = 1,\dots,D-1$ and we use the metric signature $(+1,-1,\dots,-1)$ convention.  We mention that the canonical structure of this model has been analyzed in detail for instance in references
\cite{Maharana:1983cs, Hong:2002vr}.  The Lagrangian density (\ref{ONL}) falls exactly within the scope of the prototyical system (\ref{L}), if we consider the latter written in DeWitt notation in which the repeated index summations are generalized also to continuous space integrations.   In this case, we have $M=1$,
\begin{equation}
V=\frac{1}{2p}\int d^{(D-1)} x \, \partial_i\phi^a \partial_i\phi^a\,,
\end{equation}
and the true constraints corresponding to equation (\ref{Xi1Xi2})  are given by
\begin{equation}\label{true1}
\chi_1 = T\equiv \frac{1}{2}(\phi^a\phi^a-q^2) 
\end{equation}
and
\begin{equation}\label{true2}
\chi_2 = \chi \equiv p \phi^a\pi^a 
\,,
\end{equation}
with $\pi^a$  denoting the canonical momenta conjugated to $\phi^a$.  Hence, we have
\begin{equation}
\left\{ T, \chi \right\}
= p\phi^a\phi^a \equiv w
\end{equation}
and we may read the second-class Hamiltonian directly from (\ref{Hsc}) as
\begin{equation}
H_{sc} = 
\int d^{(D-1)}x \left\{
\frac{p}{2}
\pi^a\pi^a
+\frac{1}{2p}\partial_i\phi^a\partial_i\phi^a
+\frac{p\pi^a\pi^a+p^{-1}\phi^a\nabla^2\phi^a}{2\phi^c\phi^c}\left(\phi^b\phi^b-q^2\right)
\right\}
\,,
\end{equation}
satisfying
\begin{equation}
 \left\{ H_{sc}, T \right\}
=  -\frac{p\phi^a\pi^a}{\phi^b\phi^b} T  -\chi
\end{equation}
and
\begin{equation}
 \left\{ H_{sc}, \chi \right\}
= -\frac{2p^2\phi^a\pi^a}{\phi^b\phi^b} T 
\,,
\end{equation}
corresponding to equations (\ref{SCS3}) and (\ref{SCS4}). 

The effort spent in the last sections careful analysis pays well in the current application.
Associated to this consistent description of the nonlinear sigma model, the general results from the implementation of the GU formalism in terms of our prototypical second-class system straightforwardly produce a corresponding first-class version.  Indeed, the deformed variables in phase space can be immediately obtained from (\ref{deformation}) as
\begin{equation}
\tilde{\phi}^a = \phi^a\,,~~~~\tilde{\pi}^a = \pi^a - \frac{\phi^a\phi^b\pi^b}{\phi^c\phi^c}
\,, 
\end{equation}
with the projection operator in (\ref{O}) given by 
\begin{equation}
{\cal O}^{ab}\equiv \delta^{ab}-\frac{\phi^a\phi^b}{\phi^c\phi^c}\,,
\end{equation}
and the first-class Hamiltonian (\ref{Hfc}) corresponds to
\begin{eqnarray}
H_{fc}&=&\int d^{(D-1)}x \left\{
\frac{p}{2}
\pi^a\pi^a
-
\frac{p{(\phi^a\pi^a)}^2}{2\phi^b\phi^b}
+\frac{1}{2p}\partial_i\phi^a\partial_i\phi^a
+\right.\nonumber\\&&\left.
~~~+~
\frac{p(\pi^a\pi^a - {(\phi^a\pi^a)}^2/(\phi^c\phi^c))
+p^{-1}\phi^a\nabla^2\phi^a}{2\phi^c\phi^c}\left(\phi^b\phi^b-q^2\right)
\right\} 
\,.
\end{eqnarray}
It can be checked that $H_{fc}$ is in strong involution with the first class constraints (\ref{true1}) and (\ref{true2}).  Therefore, 
we have succeeded in obtaining a nice first-class description for the nonlinear sigma model based on the GU formalism.

\section{Conclusion}
We have briefly reviewed the key aspects of the GU formalism in its improved version and contextualized it in terms of a fairly broad second-class system.  The prototypical system discussed here can be used to describe many different situations, related both to mechanical and field theory models. 
By considering the GU improved version, we have been able to obtain the general expressions for the deformed variables and corresponding first-class Hamiltonian function for the parent prototypical second-class system.  
On the one hand, it has been interesting to see that, from the Dirac-Bergmann viewpoint the prototypical system (\ref{L}) has a total of $4M$ constraints in phase space consistently conserved along time evolution.  However, on the other hand, that $4M$ constraints interpretation relies upon viewing $l^\alpha$ as independent coordinates and leads to an unnecessary proliferation of artificial constraints.  We have shown that this can be avoided by the more cleaner path chosen here, in which we identified the true constraints interrelating the physical variables. 
For a relevant physical application in field theory, we have shown how our formalism works on the nonlinear sigma model, producing a first-class Hamiltonian in strong involution with a set of first class constraints.
Other applications to continuous systems, as well as possible extensions of the prototypical second-class system, are currently under analysis.

\subsection*{Acknowledgments}
Jorge Ananias Neto thanks CNPq (Conselho Nacional de Desenvolvimento Cient\'ifico e Tecnol\'ogico), Brazilian scientific support federal agency, for partial financial support, CNPq-PQ, Grant number 307153/2020-7.

\end{document}